# Non-magnetic tight binding disorder effects in the γ sheet of *Sr₂RuO₄*.

P. Contreras[1, *], D. Osorio[1] and S. Ramazanov[2]

[1] Department of Physics, University of the Andes, Mérida, Venezuela.
[2] Department of Physics, Dagestan State University, Makhachkala, Russia.
[*]Corresponding author (pcontreras@ula.ve)

**ABSTRACT**

Inspired by the physics of the Miyake - Narikiyo model (MN) for superconductivity in the γ sheet of Sr2RuO4, we set out to investigate numerically the behavior caused by a non-magnetic disorder in the imaginary part of the elastic scattering matrix for an anisotropic tight-binding model. We perform simulations by going from the Unitary to the Born scattering limit, varying the parameter c which is inverse to the strength of the impurity potential. It is found that the unitary and intermedia limits persist for different orders of magnitude in simulating the disorder concentration. Subsequently and in order to find the MN tiny gap, we perform a numerical study of the unitary limit as a function of disorder concentration, to find the tiny anomalous gap

**Keywords**: triplet reversal time broken state, γ sheet, strontium ruthenate, non-magnetic disorder, elastic scattering matrix. Tiny Gap.

## *1. Introduction*

Strontium ruthenate (Sr$_2$RuO$_4$) [1] is a body-centered tetragonal crystal with a layered square structure for the ruthenium atoms. Its normal state is described by a Fermi liquid, with three metallic conduction sheets in the Fermi surface (FS), namely the *α*, *β*, and *γ* sheets [2]. In addition, Sr$_2$RuO$_4$ is an unconventional superconductor with $T_c \approx$ 1.5 K that strongly depends on non-magnetic disorder [3]. From the beginning, it was proposed that Sr$_2$RuO$_4$ is an unconventional superconductor with triplet pairing and some type of nodes in the order parameter for each sheet of the FS [4], where the symmetry of the superconducting gap is believed to break time reversal symmetry [5,6,7].

Although some authors consider that the γ sheet of the FS does not have nodes, several low temperature works have provided experimental agreement with nodes in the specific heat C(T), the electronic heat transport κ$_i$(T) and directional ultrasound α$_j$(T) measurements. These measurements resemble some kind of nodes in the on the FS including the $\gamma$ sheet (see [8] and references therein for a review of the experiments with the first crystal samples of Sr$_2$RuO$_4$). Recently, novel experimental and theoretical advances continue to be carried out in the comprehension of the broken time-reversal symmetry superconducting state of Sr$_2$RuO$_4$. All recent and old studies continue to be crucial in order to explain the microscopic mechanism inherent to superconductivity in this compound ( [9, 10, 11, 12] and references there in).

In this work, we use a tight binding nearest neighbor expression for $\xi_\gamma(k_x, k_y)$ in order to model the γ sheet, which is centered at (0,0) in the first Brillouin zone, $\xi_\gamma(k_x, k_y) = -\epsilon + 2\,t\,[\cos(k_x\,a) + \cos(k_y\,a)]$ with hopping parameters (t,ϵ) = (0.4, 0.4) meV, and electron-hole symmetry. For the ***k*** dependence of the gap, we use the 2D tight binding expression corresponding to the Miyake -

Narikiyo model [13] for a triplet state in the $\gamma$ sheet, that is, $\boldsymbol{\Delta}^\gamma(k_x, k_y) = \Delta_0 \, \boldsymbol{d}^\gamma(k_x, k_y)$, with the vector $\boldsymbol{d}^\gamma(k_x, k_y) = [(\sin(k_x a) + i \sin(k_y a)]\hat{\boldsymbol{z}}$ and $\Delta_0^\gamma$ = 1.0 meV, since there are experimental reports which estimated the value of $\Delta_0^\gamma$ to be less than 1 meV in impurity samples [3] (the issue of tuning $\Delta_0^\gamma$ will be considered separately).

The nine points where the order parameter (OP) $\boldsymbol{d}^\gamma(k_x, k_y)$ has zeros are sketched in fig 1, that is, 4 points symmetrically distributed in the {10} and {01} planes at k-points (0,±π) and (±π,0), 4 points symmetrically distributed in the {11} planes at k-points (±π,±π), and 1 point in the {00} plane at k-point (0,0). As noticed in [13], the gap on the $\gamma$ sheet is very anisotropic and leaves a tiny gap around (0,±π) and (±π,0) points. According to group theory considerations, the imaginary OP has two components which belong to the irreducible representation E$_{2u}$ of the tetragonal point group D$_{4h}$. It corresponds to a triplet odd paired state $\boldsymbol{d}^\gamma(-k_x, -k_y) = -\boldsymbol{d}^\gamma(k_x, k_y)$ with basis functions $\sin(k_x a)$ and $\sin(k_y a)$ and Ginzburg-Landau coefficients $(1, i)$ [14,14a]. The 3D analogous to the MN model is the Zhitomriski and Rice (ZR) interband model [16], however the ZR model has nodes on the 3 sheets of the FS, contrary to the MN model, since the OP zeros do not touch the $\gamma$ sheet, but are closed to it (tiny gap model).

In the following sections, we briefly report on a visual numerical analysis of the MN model for the $\gamma$ sheet on Sr$_2$RuO$_4$ at the phenomenological level (the microscopy mechanism is explained in their work [13]) We use a particular methodology [15]. First we vary the inverse strength parameter c from 0 to 1 and second, we vary the value of the parameter $\Gamma^+$ from optimal to dilute doping in the function $\widetilde{\omega}(\omega + i\, 0^+)$. We finish our short report by comparing our findings with a line nodes tight binding analysis for High T$_c$ compounds recently investigated using the same approach [17].

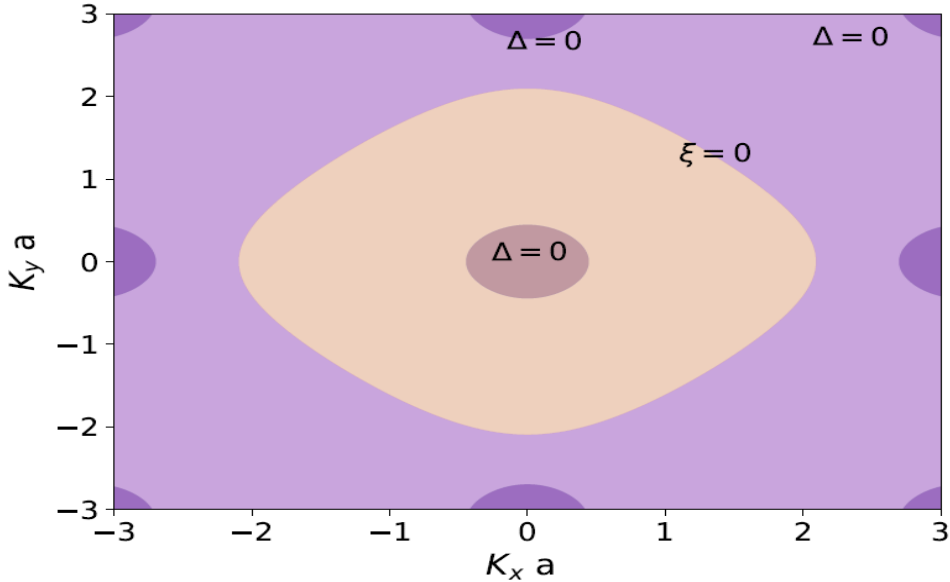

Figure 1. 2D implicit plot of the tight binding anisotropic Fermi $\gamma$ sheet $\xi_\gamma(k_x, k_y) = 0$ and the triplet superconducting gap with the localization of the nine points where $\boldsymbol{d}^\gamma(k_x, k_y) = 0$.

## 2. *From the Unitary to the Born limit in triplet superconductors*

We introduce the main equation for the elastic scattering involving the self-energy $\widetilde{\omega}(\omega + i\, 0^+)$ in the case of non-magnetic disorder according to [15] by following their approach to model experimental low temperature data in the unitary region, where the hydrodynamic limit does not work. The dressed $\widetilde{\omega}(\omega + i\, 0^+)$ can be written in the following way

$$\widetilde{\omega}(\omega + i\, 0^+) = \omega + i\,\pi\,\Gamma^+ \frac{g(\widetilde{\omega})}{c^2 + g^2(\widetilde{\omega})} \quad (1)$$

Eq. 1 describes the self-consistent renormalization in energy due to elastic scattering of superconducting pairs on non-magnetic atoms for the case of electron-hole symmetry in unconventional superconductors, i.e. $g(\widetilde{\omega}) = g_0^\gamma(\widetilde{\omega})$ (we omit the $\gamma$ label in the main equations). The parameter $c = 1/(\pi\, N_F\, U_0)$ is the inverse of the impurities strength, $N_F$ is the Fermi level DOS, and $U_0$ is the impurity potential.

The parameter $\Gamma^+ = n_{imp}/(\pi^2\, N_F)$ is proportional to the impurity concentration $n_{imp}$. Non-magnetic disorder assumes N equal scatters randomly distributed, but independent each other, it also assumes that on a macroscopic scale, the crystal is homogeneous [18]. The effect of non-magnetic disorder in $Sr_2RuO_4$ is to suppress superconductivity states around nodal/cuasi-nodal regions. The function $g(\widetilde{\omega})$ in (1) is given by $g(\widetilde{\omega}) = \langle \frac{\widetilde{\omega}}{\sqrt{\widetilde{\omega}^2 - |\Delta|^2(k_x, k_y)}} \rangle_{FS}$, and the average over the $\gamma$ sheet of the FS "$\langle \ldots \rangle_{FS}$" is performed with the tight binding expressions mentioned in the introduction, and according a numerical technique successfully used to fit experimental low temperatures ultrasound data with an accidental 3D point nodes model similar to the ZR model [19,19a,19b].

Finally, if electron-hole symmetry is not considered in the $\gamma$ sheet, the other spin Pauli components of $g^\gamma(\widetilde{\omega})$ have to be taken into consideration, that is, $g_1^\gamma(\widetilde{\omega})$ and $g_3^\gamma(\widetilde{\omega})$, but for a p-wave triplet gap $g_1^\gamma(\widetilde{\omega}) \sim \langle d^\gamma(k_x, k_y) \rangle_{FS} = 0$, and $g_3^\gamma(\widetilde{\omega}) \sim \langle \xi_\gamma(k_x, k_y) \rangle_{FS} = 0$. Born's approximation applies and if c >> 1 (i.e. $U_0$ << 1) with a new disorder doping parameter $\Gamma_B^+ = \Gamma^+/c^2$ << 1, however we do not expect Born scattering to play a role in the low temperature properties in this compound for the $\gamma$ sheet due to the fact that, any small amount of non-magnetic dirt can be in principle provided by local Sr atoms, and will cause the $\gamma$ sheet of $Sr_2RuO_4$ to be in the unitary limit, or to be close to it.

The imaginary part of (1) defines the inverse of the quasiparticle disordered averaged lifetime $\tau^{-1}(\omega)$ as

$$\tau^{-1}(\omega) = 2\,\Im\,[\widetilde{\omega}(\omega + i\,0^+)] \quad (2)$$

In the unitary limit, $\tau^{-1}(\omega)$ has a resonance at zero frecuency, that is, $\widetilde{\omega}(0) = i\,\gamma$, where $\gamma$ defines the "impurity averaged" zero energy elastic scattering rate [15], and determines the crossover energy scale separating the two scattering limits. Since we expect normal state excitations in energy (provided by very small amounts of non-magnetic local Sr atoms) to be less than $\gamma$, self-consistency in eq. [1] cannot be neglected. Finally we refer the lectors to [20] for a extended treatment of the theoretical formalism for non magnetic impurity scattering in unconventional superconductors, and to [21] for a recent review on impurity effects and modelling techniques. The impurity effect with unitarity scattering was also investigated in [22,23,24, 25, 26, 27], Finally, for Fermi and Bose atomic gases at ultra-cold temperatures in the unitary limit see [28].

If instead of investigating the tinny gap region, we set up the region where the gap Δ in magnitude is bigger than frecuency ω, the main effect in $\widetilde{\omega}(\omega + i\, 0^+)$ is to shift the real part by an amount proportional to $\omega - \Gamma^+\Delta/\widetilde{\omega}$, which tends to zero as the dressed frecuency value $\widetilde{\omega}$ is increased.

In this section, we compute the solution of eq. 1 by varying the parameter inverse to the strength c, and by fixing the value of disorder concentration $\Gamma^+$ for two cases of physical interest. The first case is for a value of $\Gamma^+ = 0.30\ meV$, which resembles optimally doped values of impurities in experimental samples. The second case is for a dilute disorder concentration $n_{imp}$, that is, $\Gamma^+ = 0.05\ meV$. In $Sr_2RuO_4$, Sr atoms in the lattice form an additional impurity level in the energy zone, thus, Sr atoms are part of the structure and also are the centers on which non-magnetic elastic scattering occurs.

Figure 2 shows the evolution of $\Im[\widetilde{\omega}](\omega + i\, 0^+)$ as function of the parameter c which is inversely proportional to the strength $U_0$, going from strong $U_o \gg 1\ (c \xrightarrow{+} 0)$ (violet curve) to values $U_o \ll 1\ c \xrightarrow{-} 1$ (yellow curve) i. e. when $\pi\, N_F\, U_0 \to 1$, $(N_F \sim 10^{15} eV^{-1})$, for an electron-hole symmetric tight binding dispersion and for an optimal disorder concentration $\Gamma^+ = 0.30$ meV.

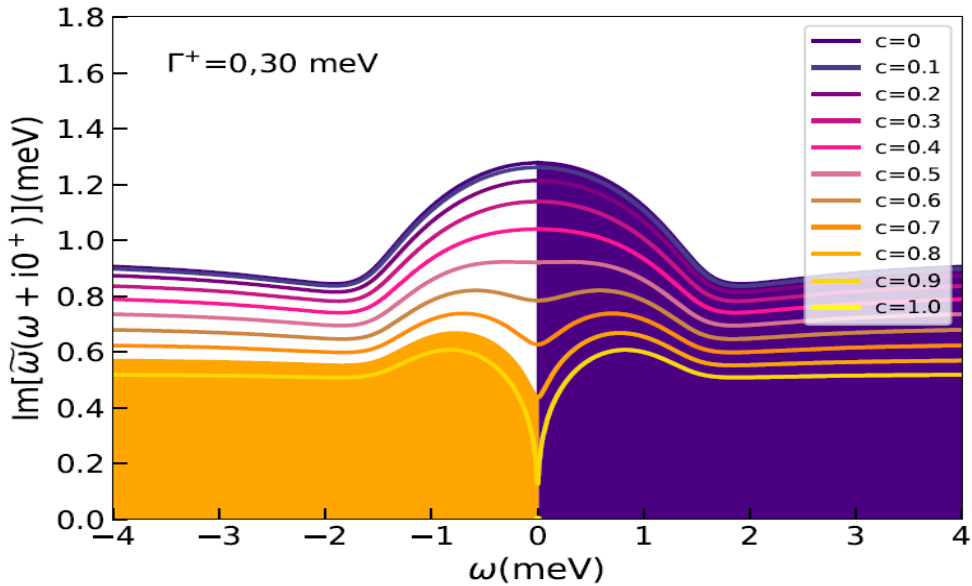

*Figure 2. Evolution of the imaginary part of the scattering matrix from the unitary limit to intermedite regions for a value Γ⁺=0.30 meV.*

The region under $\Im[\widetilde{\omega}](\omega + i\, 0^+)$ in the unitary limit which corresponds to a resonance at zero frequencie has been shaded violet in the right side of the plot. On the other hand, a well define intermedia scattering region with a finite minimum at zero frequencies, and resonances at ω ∼ ± 0.8 meV, wich correspond to an inverse strength of c = 1 has been shade orange in the left side.

We observe the unitary behavior from c = 0 to c = 0.5, and well defined intermedia regions from c =

0.6 up to values c = 1, that is, $U_0 \to \pi^{-1} N_F^{-1}$. In figure 2, non-magnetic disorder affects most strongly the low energy region up to 1.5 meV. We also see from fig. 2 that in the normal state, for energies bigger that the gap value, the function $\Im[\widetilde{\omega}](\omega + i\, 0^+)$ becomes constant, but tends to depend on the c value.

Figure 3 shows the evolution of $\Im[\widetilde{\omega}](\omega + i\, 0^+)$ as a function of c for a dilute concentration of impurities, that is, $\Gamma^+ = 0.05$ meV. The c = 0.1 region for $\Im[\widetilde{\omega}](\omega + i\, 0^+)$ in the unitary limit has been shaded blue in the right side of the plot, we observe a much smaller "impurity averaged" zero energy elastic scattering rate ($\gamma \sim 0.4\, meV$) that in the previous case ($\gamma \sim 1.3\, meV$).

On the other hand, the intermedia scattering region ($\gamma \sim 0.05\, meV$) which corresponds to a value of c = 0.4 has been shaded rose in the left side of fig 3. We observe a flattening of the function for higher energies than 1.7 meV, which indicates a constant normal state quasiparticle for the case of dilute impurity concentration.

Finally we observe in figure 3, an anomalous drop to zero of the function $\Im[\widetilde{\omega}](\omega + i\, 0^+)$ around $\omega \sim 1\, meV$ for the five values of c where the solution was found. Therefore, in the next section we study the unitary limit c = 0 in order to further investigate into that anomaly.

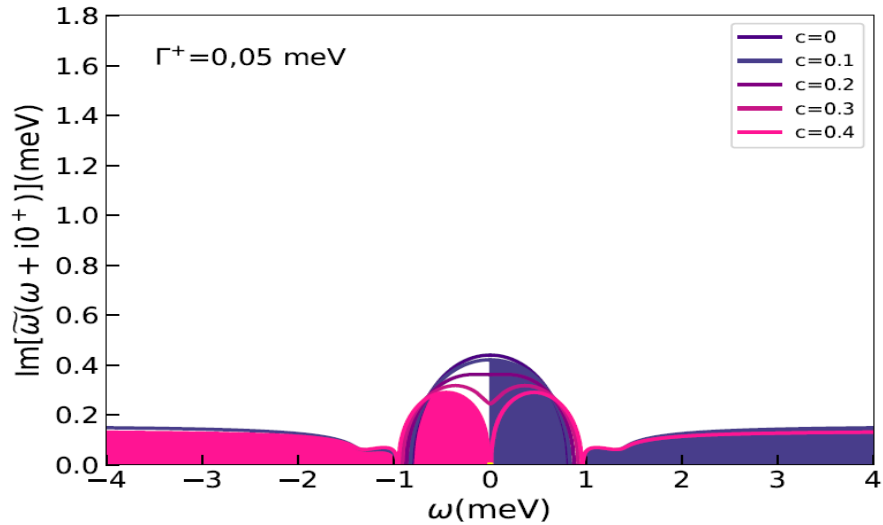

*Figure 3. Evolution of the imaginary part of the scattering matrix from the unitary limit to intermedite regions for a value Γ⁺=0.05 meV.*

### 3. Inside the unitary limit, the visualization of the Miyake Narikiyo tinny gap in the $\gamma$ sheet

In this section, we visualize the behavior of $\Im[\widetilde{\omega}](\omega + i\, 0^+)$ inside the unitary limit (c = 0) for different values of the impurities concentration parameter $\Gamma^+$, starting at very dilute disorder (yellow line), to an optimal disorder (violet line). For large values of $U_0$, the unitary limit in eq. 1 is given by expression

$$\widetilde{\omega}(\omega + i\, 0^+) = \omega + i\, \pi\, \Gamma^+ \frac{1}{g(\widetilde{\omega})} \quad (3)$$

The unitary regime is defined as the limit where the elastic scattering due to non-magnetic disorder is so strong that the mean-free path becomes comparable to the Fermi momentum $k_F$, meanwhile in the Born metallic limit, the mean free path is much larger than the Fermi momentum $k_F$. This means that normal state quasiparticles in the unitary region have an ill-defined momentum between elastic collisions. In this formalism, the signature of the unitary state is the resonance at zero frequency in the imaginary part of the scattering matrix.

Figure 4 shows the evolution of $\Im[\widetilde{\omega}](\omega + i\,0^+)$ according to eq. 3, and for nine values of $\Gamma^+$. We observe the smooth resonance centered at zero frequency for all values, with smaller values of residual zero energy $\gamma$ for very dilute values of disorder $\Gamma^+$. The region for $\Im[\widetilde{\omega}](\omega + i\,0^+)$ corresponding to an optimal levels of disorder, $\Gamma^+ = 0.40$ meV, has been shaded violet in the right side of the plot. In addition, the region corresponding to dilute levels of disorder with $\Gamma^+ = 0.05$ meV, has been shaded yellow in the left side of the figure.

In the left part of the fig. 4, the MN tiny gap, that was numerically observed by using a density of states DOS analysis, is also found in the $\Im[\widetilde{\omega}](\omega + i\,0^+)$ analysis as well. The tiny gap is given by $\Im[\widetilde{\omega}](\omega + i\,0^+) = 0$ and is found in the interval between 0.85 meV and 1 meV, corresponding to a 15 % of the value $\Delta_0 = 1$ meV used. This tiny gap effect in the function $\Im[\widetilde{\omega}(\omega + i\,0^+)] \sim \tau^{-1}(\omega)$ will have significance for the low temperature transport properties since it enters the expressions for several kinetic coefficients.

In order to compare directly with, we performed an analysis using the same methodology for a lines nodes tight binding model for a High $T_c$ materials [16]. It allows us to state, that the case of the $\gamma$ sheet tight binding analysis on $Sr_2RuO_4$, is much more sensitive to strong elastic scattering events and self consistency at low temperatures, since the unitary limit persists for most of the values used for the modeling parameters c and $\Gamma^+$.

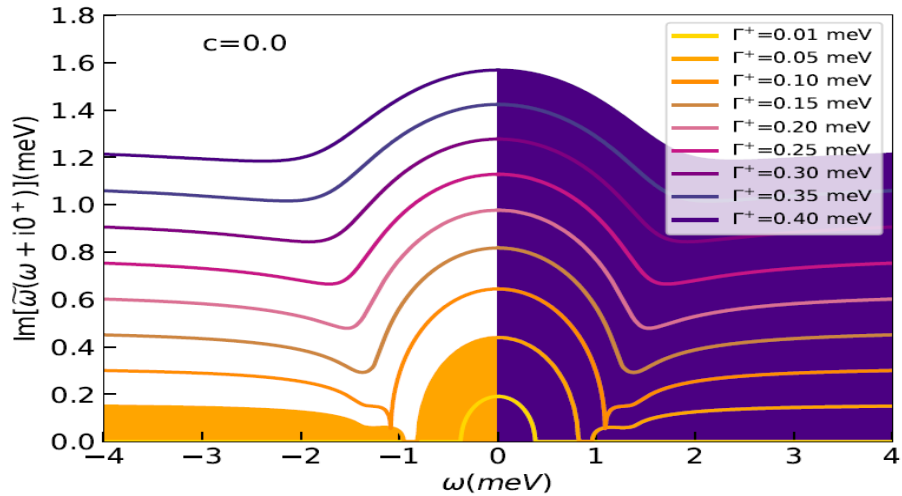

*Figure 4. Evolution of the imaginary part of the scattering matrix inside the unitary limit for nine values of the parameter $\Gamma^+$(meV).*

## 4. Conclusion

This communication was aimed at investigating numerically the behavior of the elastic scattering non-magnetic disordered averaged matrix $\widetilde{\omega}(\omega + i\, 0^+)$ for a Miyake Narikiyo 2D anisotropic tigh binding model for the γ sheet of $Sr_2RuO_4$. In section 2, we modeled and visualized the behavior of the imaginary part $\mathfrak{I}[\widetilde{\omega}](\omega + i\, 0^+)$ of the scattering matrix. The dependence on the inverse of the strength of the disorder potential $U_0$, i.e. the parameter c, was studied for two regions of physical importance, an optimal disorder region with $\Gamma^+$ = 0.40 meV, and a dilute region with $\Gamma^+$ = 0.05 meV. The results were illustrated in figs. 2 and 3.

We found that the function $\mathfrak{I}[\widetilde{\omega}](\omega + i\, 0^+)$ is always within the unitary or intermedia scattering limits for the values of the parameters used, contrasting with the case of a High Tc modeling where the Born hydrodynamic limit is present at small values of the parameter c [16]. In section 3, the behavior of the disordered matrix $\widetilde{\omega}(\omega + i\, 0^+)$ inside the unitary (c = 0) region was studied for nine values of $\Gamma^+$, starting at very diluted disorder, optimal disorder values, and finally an enriched disorder.

The results were visualized in figure 4, the tiny Miyake – Narikiyo gap was found for a $\Gamma^+$= 0.05 meV in disorder concentration. We end this short note pointing out that the Miyake – Narikiyo model is very useful for setting up numerical studies in triplet superconductors such as strontium ruthenate, as we have demonstrated in this study.

## 5. Acknowledgements

The authors did not receive financial support for research, authorship and/or publication of this article.